\documentclass[showpacs,reprint,twocolumn,floatfix,aps,prapplied,longbibliography]{revtex4-1}

\usepackage{graphicx,float,amsmath}
\usepackage{natbib}
\usepackage{lipsum}
\usepackage{xcolor}

\begin{document}

\title{Piezoresistance in defect-engineered silicon}

\author{H. Li$^1$}
\author{A. Thayil$^1$}
\author{C.T.K. Lew$^2$}
\author{M. Filoche$^1$}
\author{B.C. Johnson$^2$}
\author{J.C. McCallum$^3$}
\author{S. Arscott$^4$}
\author{A.C.H. Rowe$^1$}
\email{alistair.rowe@polytechnique.edu}
\affiliation{$^1$Laboratoire de Physique de la Mati\`ere Condens\'ee, Ecole Polytechnique, CNRS, IP Paris, 91128 Palaiseau, France}
\affiliation{$^2$Centre for Quantum Computation \& Communication Technology, School of Physics, University of Melbourne, VIC 3010, Australia}
\affiliation{$^3$School of Physics, University of Melbourne, Melbourne, Victoria 3010, Australia  }
\affiliation{$^4$Institut d'Electronique, de Micro\'electronique et de Nanotechnologie (IEMN), Universit\'e de Lille 1, CNRS, Avenue Poincar\'e, Cit\'e Scientifique, 59652 Villeneuve d'Ascq, France}

\begin{abstract}

The steady-state\textcolor{red}{, space-charge-limited}  piezoresistance (PZR) of defect-engineered, silicon-on-insulator device layers containing silicon divacancy defects changes sign as a function of applied bias. Above a punch-through voltage ($V_t$) corresponding to the onset of a space-charge-limited hole current, the longitudinal $\langle 110 \rangle$ PZR $\pi$-coefficient is $\pi \approx 65 \times 10^{-11}$~Pa$^{-1}$, similar to the value obtained in charge-neutral, p-type silicon. Below $V_t$, the mechanical stress dependence of the Shockley-Read-Hall (SRH) recombination parameters, specifically the divacancy trap energy $E_T$ which is estimated to vary by $\approx 30$~$\mu$V/MPa, yields $\pi \approx -25 \times 10^{-11}$~Pa$^{-1}$. \textcolor{red}{The combination of space-charge-limited transport and defect engineering which significantly reduces SRH recombination lifetimes makes} this work \textcolor{red}{directly} relevant to discussions of giant or anomalous PZR \textcolor{red}{at small strains} in \textcolor{red}{nano-}silicon \textcolor{red}{whose characteristic dimension is larger than a few nanometers. In this limit the} reduced \textcolor{red}{electrostatic} dimensionality lowers $V_t$ and amplifies space-charge-limited currents and efficient SRH recombination occurs via surface defects. \textcolor{red}{The results} reinforce the growing evidence that in steady state, electro-mechanically active defects can result in anomalous, but not giant, PZR.

\end{abstract}
\maketitle

\section{Introduction}
\textcolor{red}{The effect of mechanical stress in nano-silicon has received significant attention over the last two decades. Initially this was triggered by the observation of large, stress-induced mobility increases in quantum confined inversion layers \cite{dorda1971,welser1994,fischetti2002,fischetti2003} that, along with subsequent work \cite{thompson2006,sun2007}, lead to commercialized, strained-silicon CMOS technologies. More recent developments in computing power allowed for a large number of atomistic theoretical studies on a variety of ultra-quantum-confined nano-silicon with a characteristic dimension below approximately 4 nm \cite{kim2011}. These works predict a number of intriguing electronic structure and transport phenomena, particularly at very large mechanical strains exceeding 2 \% \cite{hong2008,leu2008,wu2009}. Many of these predictions remain to be validated experimentally, partly because the fabrication of such small, electrically contacted nanostructures with well-controlled surfaces is a challenge, and partly because the application of such large, non-destructive stresses is not straight-forward. With very few exceptions \cite{lugstein2010,kumar2013}, experimental works reported to date treat nano-silicon objects such as nanowires and nanomembranes whose characteristic dimension lies between several tens of nanometres and a few microns, or where mechanical strains fall in the 0.01 \% range. Although their electronic structure is simply that of bulk silicon, there are multiple claims and observations of either giant \cite{he2006,reck2008,neuzil2010,kang2012} or anomalous \cite{lugstein2010,jang2014,winkler2015} piezoresistance (PZR) that are significantly different from the usual effect observed in bulk material \cite{smith1954}. These nanostructures are too large for the types of phenomena predicted to occur in ultra-quantum-confined nano-silicon \cite{cao2007}, and there is as yet no satisfactory physical explanation of these effects. Indeed, in some cases even the veracity of the observations is contested \cite{rowe2014}. It is this large nanostructure, small strain limit which is addressed here.}

\textcolor{red}{After the initial report of giant PZR in suspended silicon nanowires \cite{he2006}, it was rapidly realized that large stress-induced resistance changes were correlated with equilibrium carrier depletion \cite{rowe2008}, and that electronically active defects, possibly at the surface, have some role to play \cite{rowe2008,reck2008,yang2010,yang2011,kang2012,jang2014,winkler2015}.  
In parallel with these reports, it was shown that under conditions of carrier depletion, large non-stress-related drifts in device currents are possible, and that these may be easily confused with unusual PZR if care is not taken to separate them, for example by modulating the applied mechanical stress \cite{milne2010}. Studies where such precautions are taken generally find that nano-silicon exhibits either the usual bulk silicon PZR \cite{mile2010,koumela2011,bosseboeuf2015} or anomalous (but not giant) PZR where the sign changes relative to that expected for the given doping type \cite{yang2010,yang2011,jang2014,winkler2015}. Given this, it is reasonable to ask why anomalous PZR is only sometimes reported, why giant PZR is so elusive, why carrier depletion is important, and what the role of electrically active defects is.}

\textcolor{red}{To investigate these questions further,} here the role of electrically-active defects and partial charge carrier depletion is made explicit by deliberately introducing silicon divacancy defects into thin device layers of so-called fully-depleted silicon-on-insulator (SOI) via self-implantation of Si$^{5+}$ ions. \textcolor{red}{Using a numerical solution of the stress-dependent, coupled Poisson and charge transport equations to simulate the transport,} an observed, anomalous sign change of the PZR as a function of the applied bias is quantitatively attributed to a combination of the bipolar nature of \textcolor{red}{space-charge-limited currents (SCLC), and to the stress dependence of the divacancy trap energies that modifies the Shockley-Read-Hall (SRH) recombination rate.}

PZR has historically been studied in doped, bulk semiconductor devices at low applied voltages where it is reasonable to assume unipolar electrical transport in the charge-neutral limit, i.e. in which the density of non-equilibrium injected charge is negligible compared to the equilibrium free charge density \cite{sze2007}. Under such conditions ohmic conduction is observed i.e. the current density, $J$, is proportional to the applied voltage, $V$. For the case of p-type material $J = \sigma_p V/d$ where $d$ is the channel length, $\sigma_p = 1/\rho_p = p\mu_pq$ is the hole conductivity, $p$ is the hole density, $\mu_p$ is the hole mobility and $q$ is the electronic charge. A similar expression can be given for electrons. The PZR in charge-neutral silicon is principally the result of mechanical-stress-induced changes to the effective masses and hence the mobilities \cite{smith1954}, and its sign is determined only by the doping type. Generally speaking, since the effective masses are tensor quantities in a crystal, so too is the PZR. However, for the case of a resistance measurement made parallel to the direction of the applied stress, the PZR is characterized by a scalar, longitudinal $\pi$-coefficient which, in the case of holes, is: \begin{equation} \label{pi} \pi_p = \frac{1}{X}\frac{\Delta \rho_p}{\rho_{p0}} \approx -\frac{1}{X}\frac{\Delta \mu_p}{\mu_{p0}}, \end{equation} where $\mu_{p0}$ is the zero-stress mobility and $X$ is the applied stress. The approximate equality is valid for small changes in the mobility. Once again, a similar expression can be given for electrons. In the devices considered here, resistance is measured parallel to an applied stress along the $\langle 110 \rangle$ crystal direction for which \cite{kanda1982} \begin{equation} \label{pip} \pi_p \approx +71 \times 10^{-11} \:\textrm{Pa}^{-1}, \end{equation} and \begin{equation} \label{pin} \pi_n \approx -30 \times 10^{-11} \:\textrm{Pa}^{-1}. \end{equation} While the steady-state PZR measured here is approximately bounded by these values, it is not \textit{only} due to stress-induced mobility changes. 

\section{Sample details}

\begin{figure}[t]
\includegraphics[clip,width=8.5 cm] {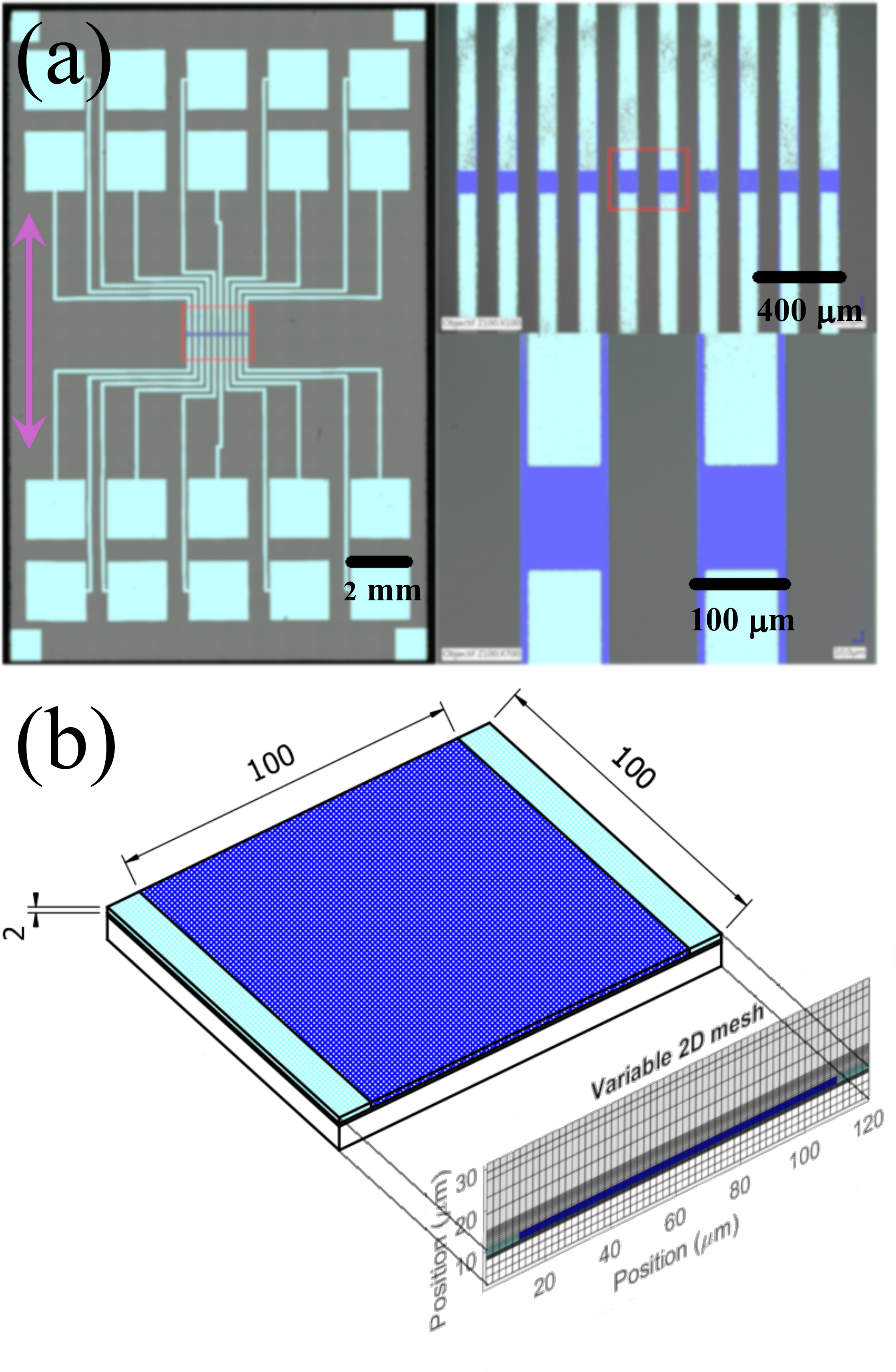}
\caption{(a) False color micrographs of the devices. The left panel shows the macroscopic chip layout with the external, metallized ohmic contacts clearly visible in light blue. As indicated by the red box, a zoom of multiple devices is shown in the top, right panel. A further zoom to the individual device level is shown in the bottom, right panel. The active device areas are shown in dark blue. Tensile mechanical stress is applied parallel to the $\langle 110 \rangle$ crystal direction as indicated by the purple arrow in the left panel. (b) A perspective schematic diagram of an individual device using the same color scheme as the micrographs, with active volume dimensions shown in $\mu$m. A variable mesh projection in the vertical plane to be used in the device modeling and analysis is also shown.}
\label{sample_details}
\end{figure}

Two-terminal devices are fabricated using standard photo-lithographic processing methods from $(001)$-oriented, fully-depleted SOI with a 2 $\mu$m-thick, non-intentionally-doped device layer (DL) shown in dark blue in Fig.~\ref{sample_details} and a 1 $\mu$m-thick buried oxide layer (BOX). Devices of the type used elsewhere \cite{milne2010,li2019} are fabricated with $p^{+}$-ohmic contacts (boron, $10^{18}$~cm$^{-3}$) shown in light blue in Fig.~\ref{sample_details}, and then cut into chips (20~mm $\times$ 13~mm) whose long axis is parallel to the $\langle 110 \rangle$ crystal direction as seen in the left panel of Fig.~\ref{sample_details}(a). These chips are compatible with a 3-point bending apparatus and approach described elsewhere \cite{milne2010,mcclarty2016,li2019} that is used here to apply a time-modulated, tensile mechanical stress of $\approx$ 20~MPa for the PZR measurements along the $\langle 110 \rangle$ crystal direction as indicated by the purple arrow in Fig.~\ref{sample_details}(a). Fig.~\ref{sample_details}(a) shows progressive zooms of the devices from the chip level in the left panel, to the multi-device level in the top, right panel, to the individual device level in the bottom, right panel. The zooms are indicated by the red rectangles in the figure. The lateral dimensions of an individual device's active area between the ohmic contacts are 100~$\mu$m $\times$ 100~$\mu$m. Fig.~\ref{sample_details}(b) shows a perspective schematic drawing of an individual device using the same color code as the micrograph images. In the perspective drawing the top 8~$\mu$m of the 400~$\mu$m thick handle is shown in white, the buried oxide (BOX) shown in dark gray, the device layer is shown in dark blue, and the $p^{+}$ contacts are shown in light blue. All dimensions are in micrometers. The variable mesh projection will be used for the device modeling and analysis, and will be commented on further below.    

\begin{figure}[t]
\includegraphics[clip,width=8 cm] {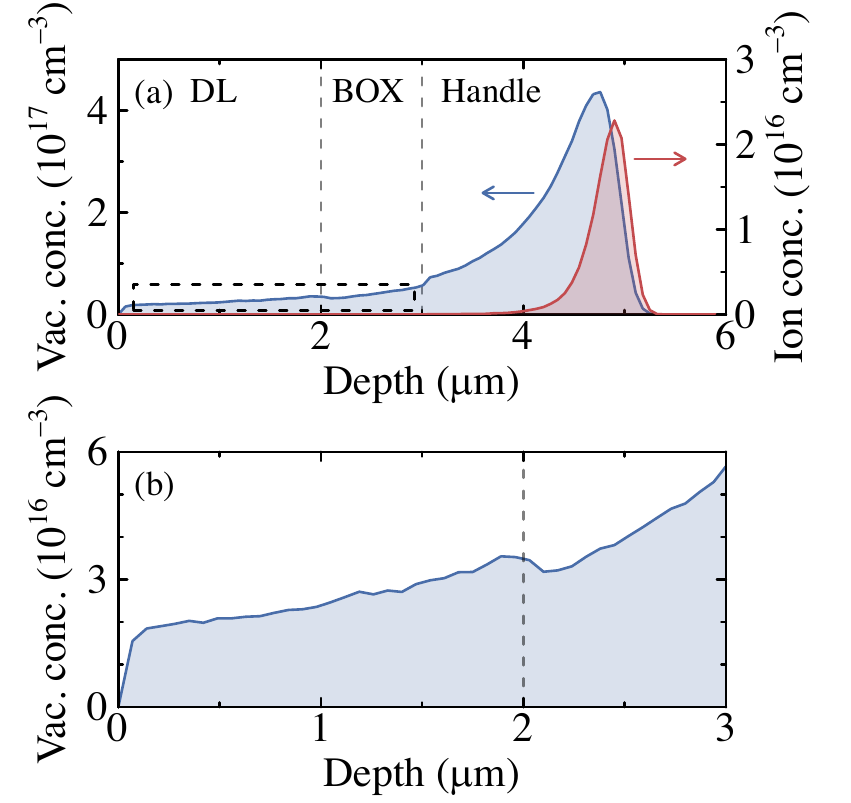}
\caption{(a) SRIM modeling of the 10 MeV Si$^{5+}$ ion implant into the silicon-on-insulator wafers. The implant principally occurs in the wafer handle (red curve in the top panel), but a long tale of implanted ions on the device layer side of the results in an approximately homogeneous distribution of silicon divacancy defects of density $\approx 2.5 \times 10^{16}$~cm$^{-3}$ in the device layer itself (black curves, including device layer zoom in (b)).}
\label{srim}
\end{figure}

Post-processing, a selection of 20~mm $\times$ 13~mm chips are exposed to a 10 MeV beam of Si$^{5+}$ ions with the aim of forming a desired density of silicon divacancy defects \cite{chason1997}. The total resulting dose is 10$^{12}$~cm$^{-2}$ which Stopping and Range of Ions in Matter (SRIM) modeling \cite{ziegler2010} indicates should result in a deposition of the majority of the ions into the wafer handle as seen in the red curve of Fig.~\ref{srim}(a). The SRIM modeling also allows for a calculation of the resulting nominal silicon divacancy defect concentration as a function of depth (blue curve in Fig.~\ref{srim}(a)). A closer inspection of the device layer itself, shown in Fig.~\ref{srim}(b), shows that this should result in an approximately uniform distribution of divacancy defects in the device layer of density $\approx 2.5 \times 10^{16}$~cm$^{-3}$.

\begin{figure}[t]
\includegraphics[clip,width=8 cm] {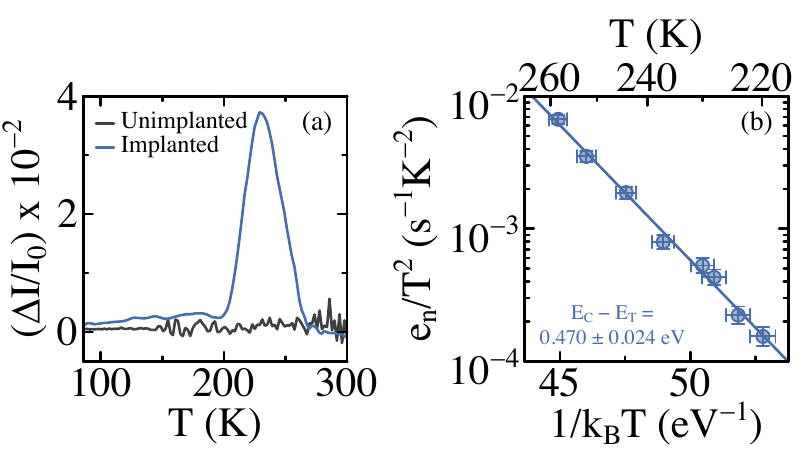}
\caption{(a) Typical PICTS signal obtained before (black curve) and after (blue curve) Si$^{5+}$ ion implantation. The implantation results in a single peak around 240~K, and a standard double box car analysis yields the Arrhenius plot shown in (b), the slope of which corresponds to a deep electronic trap 0.47~eV below the conduction band edge}
\label{picts}
\end{figure}

In order to evaluate the result of the ion implantion, photo-induced current transient spectroscopy (PICTS) \cite{papaioannou1989} on the resulting devices using a 940~nm laser with a 20~ns rise/fall time were performed using a home-built deep level transient spectroscopy (DLTS) setup. Photo-induced current transients are measured using a fast current amplifier at temperatures ranging from 80~K to 300~K with a fixed bias of 6~V applied to the samples. The resulting PICTS signal i.e. the current difference obtained using a double box car technique, shows a single peak around 240~K after defect engineering (blue curve, Fig.~\ref{picts}(a)) whereas the PICTS signal before defect engineering is featureless (black curve, Fig.~\ref{picts}(a)). Using the usual DLTS methods to obtain the temperature dependence of the emission rates from the electronic trap responsible for the PICTS peak, the Arrhenius plot in Fig.~\ref{picts}(b) is obtained. The slope yields an activation energy of 0.47~eV for the electronic trap which is therefore tentatively identified as the singly ionized acceptor form of the silicon divacancy defect \cite{svensson1991}. The absence of other defect signals, particularly the other charge states of the divacancy defect, suggests that the singly charged state is the most energetically favorable or that its optical capture rate is the fastest. Strictly speaking, this cannot however rule out the presence of other implant-induced defects in the sample.

The two principal effects of the Si$^{5+}$ irradiation are to drastically shorten the Shockley-Read-Hall (SRH) electron and hole lifetimes \cite{wright2008} and to potentially modify the type and density of the equilibrium doping density in the non-intentionally-doped active area of the device between the ohmic contacts \cite{iwata2004}. It will be seen below that the defect engineering results in a lightly, n-type active area so that the devices formed are p$^+$/n/p$^+$ bipolar structures in which the lifetime of any injected, non-equilibrium charge is orders of magnitude shorter than the lifetimes of un-irradiated silicon.

\section{Zero-stress characteristics}

\begin{figure}[t]
\includegraphics[clip,width=8 cm] {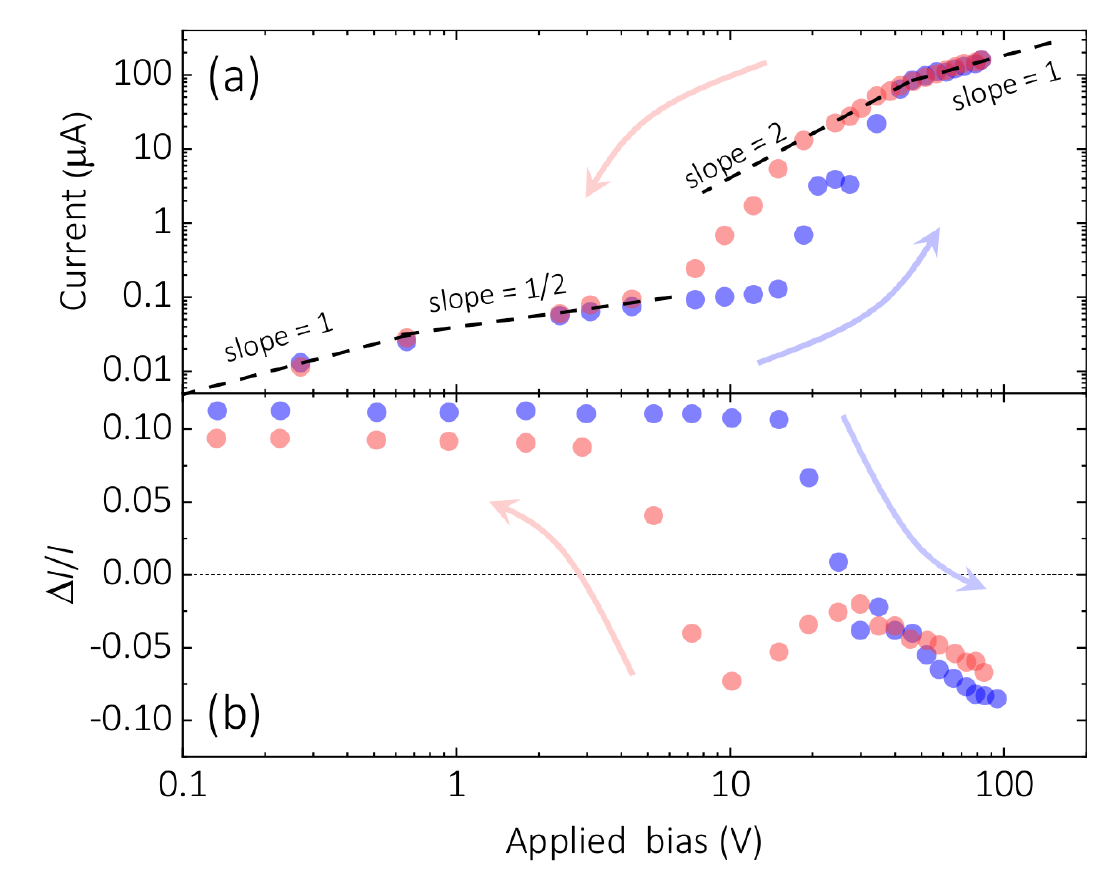}
\caption{(a) Experimentally measured up- (blue dots and arrow) and down- (red dots and arrow) sweep current-voltage characteristics obtained on the defect engineered devices. The slopes indicated in the log-log plot are a guide to the eye. (b) Relative current change induced by a +1~V change in the voltage applied to the wafer handle. The sign indicates a majority electron current below a threshold voltage, $V_t$, and a majority hole current above this bias. A hysteresis in $V_t$ is clearly visible between the up- and down-sweeps.}
\label{data}
\end{figure}

Figure \ref{data}(a) shows typical zero-stress, current-voltage characteristics obtained in a defect-engineered sample with the wafer handle held at ground. The arrows and colors represent the direction of the bias sweep, and a hysteresis is visible between the up (blue markers and arrow) and the down (red markers and arrow) sweeps. The curves were obtained in quasi-steady-state by applying a series of fixed voltage biases and then waiting until the current stabilized at each point. Stabilization times are in general of the order of a few minutes at most, except near the threshold voltage, $V_t$, where the current abruptly increases. In this bias range stabilization times are long, sometimes of the order of one day or more, and therefore the steady-state nature of the current cannot be guaranteed around $V_t$. Most importantly for this work however, is that at biases around $V_t$ the majority carrier in the active area changes from electrons to holes. The evidence for this is shown in Fig.~\ref{data}(b) which shows the relative current changes induced by a +1 V change in the handle voltage which acts as a gate for the device layer. Below $V_t$ an increase in the current indicates that electrons are the majority carriers in the active area while, on the contrary, above $V_t$ holes become the majority carrier. This is the typical behavior observed in the punch-through effect in p$^+$/n/p$^+$ bipolar junction devices \cite{lohstroh1981}. 

To better understand the \textcolor{red}{macroscopic} electrical properties of the defect engineered devices, a \textcolor{red}{self-consistent} numerical solution of the Poisson/drift-diffusion equations \textcolor{red}{in the van Roosbroeck form} is sought. \textcolor{red}{Although the drift-diffusion equations include spontaneous band-to-band recombination, SRH recombination and Auger recombination terms, at the injection levels used here it is found that the SRH process is dominant.} \textcolor{red}{The implementation follows the standard} Scharfetter-Gummel approach \cite{scharfetter1969} on a variable rectangular mesh like that shown in Fig.~\ref{sample_details}(b). The \textcolor{red}{calculation} is performed on a 2-dimensional mesh in order to properly account for the reduced electrostatic dimensionality of the devices \textcolor{red}{\cite{grinberg1989, alagha2017}} and the presence of a low permittivity environment (air), both of which affect $V_t$ \textcolor{red}{and} the magnitude of the SCLC \cite{alagha2017}. 

\begin{figure}[t]
\includegraphics[clip,width=8 cm] {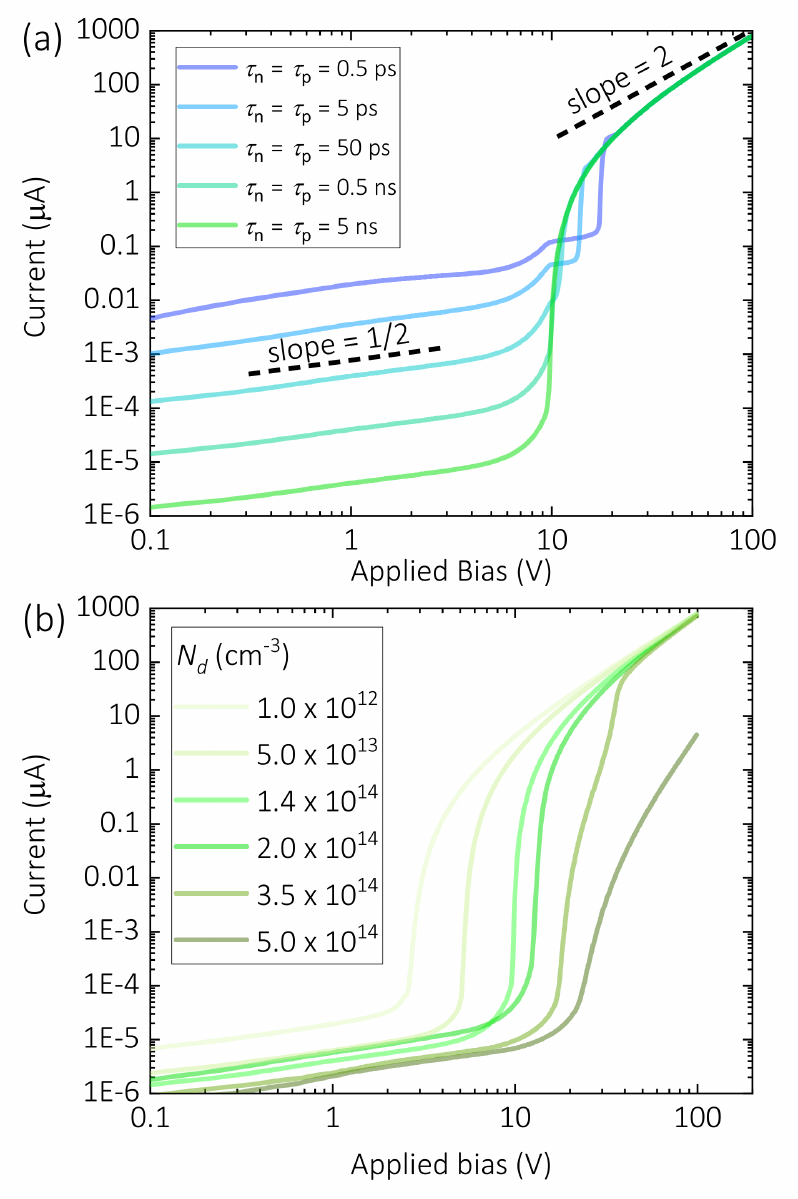}
\caption{Calculated dependence of the device characteristics on (a) the SRH recombination times, $\tau_n$ and $\tau_p$, for a donor density $N_d = 1.4\times 10^{14}$~cm$^{-3}$ and (b), on the donor density, $N_d$ for $\tau_n = \tau_p = 5$~ns. The lifetimes principally change the sub-threshold electron current which is recombination limited, and $N_d$ principally changes the threshold voltage $V_t$ corresponding to the rapid increase in current and the onset of a space-charge-limited hole current.}
\label{model}
\end{figure}

The defect engineering on the devices considered here is accounted for in the model by using drastically reduced SRH lifetimes \cite{wright2008} and by introducing a small donor density, $N_d$, presumably arising from secondary effects of the ion implantation which renders the active area n-type \cite{iwata2004}. Fig.~\ref{model} shows the principal effects of a change in these parameter values on the calculated current-voltage characteristics. Fig.~\ref{model}(a) explores the effect of a change in the SRH electron and hole lifetimes, $\tau_n$ and $\tau_p$, respectively for a donor density $N_d = 1.4 \times 10^{14}$~cm$^{-3}$. While for the shortest times (i.e. below 5 ps) there is a slight shift in the threshold voltage $V_t$, the principal effect of a reduction in the lifetimes is to increase the sub-threshold current. Note that in this sub-threshold region the current varies as $\sqrt{V}$ as expected for a recombination-limited minority current (here electrons) between two reservoirs of majority carriers (here holes). Above threshold a $V^2$-dependence typical of a Mott-Gurney like SCLC is calculated. As will be discussed below this is indeed a SCLC of holes injected from the p$^+$ contacts. Fig.~\ref{model}(b) shows the variation in the calculated characteristics for $\tau_n = \tau_p = 5$~ns (i.e. the green curve in Fig.~\ref{model}(a)) when $N_d$ is varied. While there are relatively small changes in the sub-threshold current, the principal effect of a change in $N_d$ is to change the threshold voltage $V_t$ itself. Therefore, in trying to match as best as possible the calculated characteristics with the measured data, the SRH lifetimes are first estimated from the low-voltage current and then $N_d$ is subsequently determined from $V_t$.

\begin{figure}[t]
\includegraphics[clip,width=8 cm] {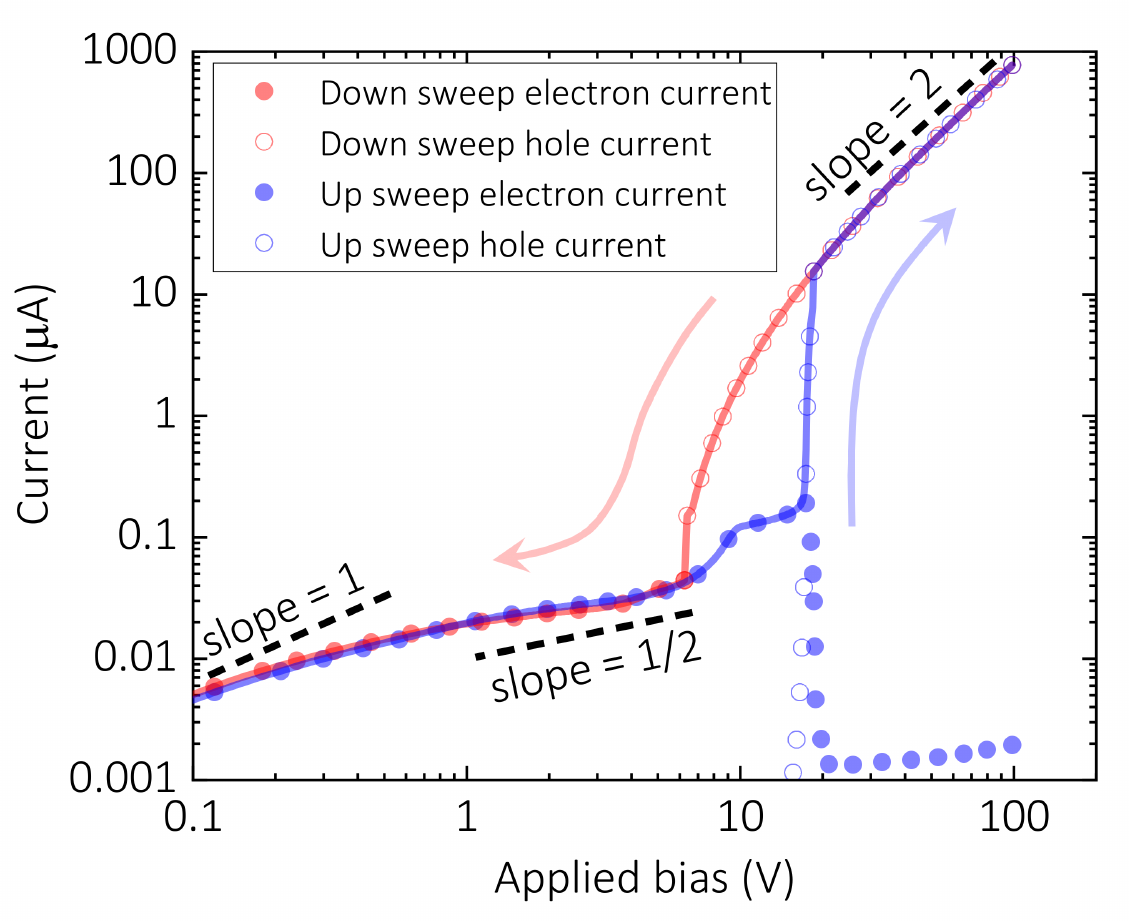}
\caption{The calculated current-voltage characteristics (solid lines) calculated using a self-consistent solution of the Poisson-transport equations \cite{scharfetter1969}. The colors correspond to the voltage sweep direction as indicated by the arrows, and the symbols illustrate the switch from an electron to a hole current at $V_t$. The model qualitatively reproduces the punch-through behavior observed experimentally in Fig.~\ref{data}}
\label{Calc_IV}
\end{figure}

Figure \ref{Calc_IV} shows the calculated current-voltage characteristics that best match the experimental data in Fig.~\ref{data}(a). Extremely short SRH carrier lifetimes of 0.5~ps consistent with values obtained after Si$^{5+}$-ion implantation \cite{wright2008} are used, and $N_d = 1.4 \times 10^{14}$~cm$^{-3}$ for the up-sweep characteristic (blue curve) while $N_d = 0.5 \times 10^{14}$~cm$^{-3}$ for the down-sweep characteristic (red curve). In both cases the upper limits for the carrier mobilities are used, $\mu_n = 1400$~cm$^2$/Vs and $\mu_p = 450$~cm$^2$/Vs. \textcolor{red}{Given the strong dependence of the magnitude of SCLCs on the geometry of the sample \cite{grinberg1989,alagha2017}, this agreement a posteriori suggests that the charge carrier mobilities are not significantly reduced via scattering \cite{dupre2007} from the engineered divacancy defects. In addition to the current magnitude,} many of the features of the experimentally measured data in Fig.~\ref{data}(a) are reproduced, including the punch-through effect in which the current is dominated by electrons below $V_t$ (filled circles) and holes above it (empty circles), and the variation from ohmic behavior at very low voltages to a $\sqrt{V}$-dependence below $V_t$. There are however some differences between the modeled and measured characteristics. For example, the experimentally observed hysteresis can be reproduced by varying $N_d$, suggesting that the application of large applied biases affects the donor charge state. While it is possible to speculate about the the details of this electric-field-activated process \cite{murgatroyd1970,ganichev2000}, in the real devices it is likely to occur progressively with applied bias. Consequently, the exact shapes of the calculated and measured current-voltage characteristics are not expected to match perfectly. Another difference occurs at high voltages where the model produces a typical $V^2$ SCLC characteristic \cite{mott1940,alagha2017} whereas the data in Fig.~\ref{data}(a) shows a linear dependence. The model includes velocity saturation so this does not account for the linear characteristic. It is likely that the linearity is due to a potential barrier at the contacts which limits hole injection \cite{rohr2018}, and which is not accounted for in the model. As will be discussed in section \ref{PZRsection}, the PZR data at high voltage support this conclusion. 

Therefore despite the excellent qualitative agreement between the model and the experimental data, the model is not expected to yield a fit to the experimental curves. It is rather aimed at aiding in the physical interpretation of the PZR data when stress-dependent quantities are introduced into the model, and in this it proves to be very useful.  

\section{Piezoresistance}
\label{PZRsection}

\begin{figure}[t]
\includegraphics[clip,width=8 cm] {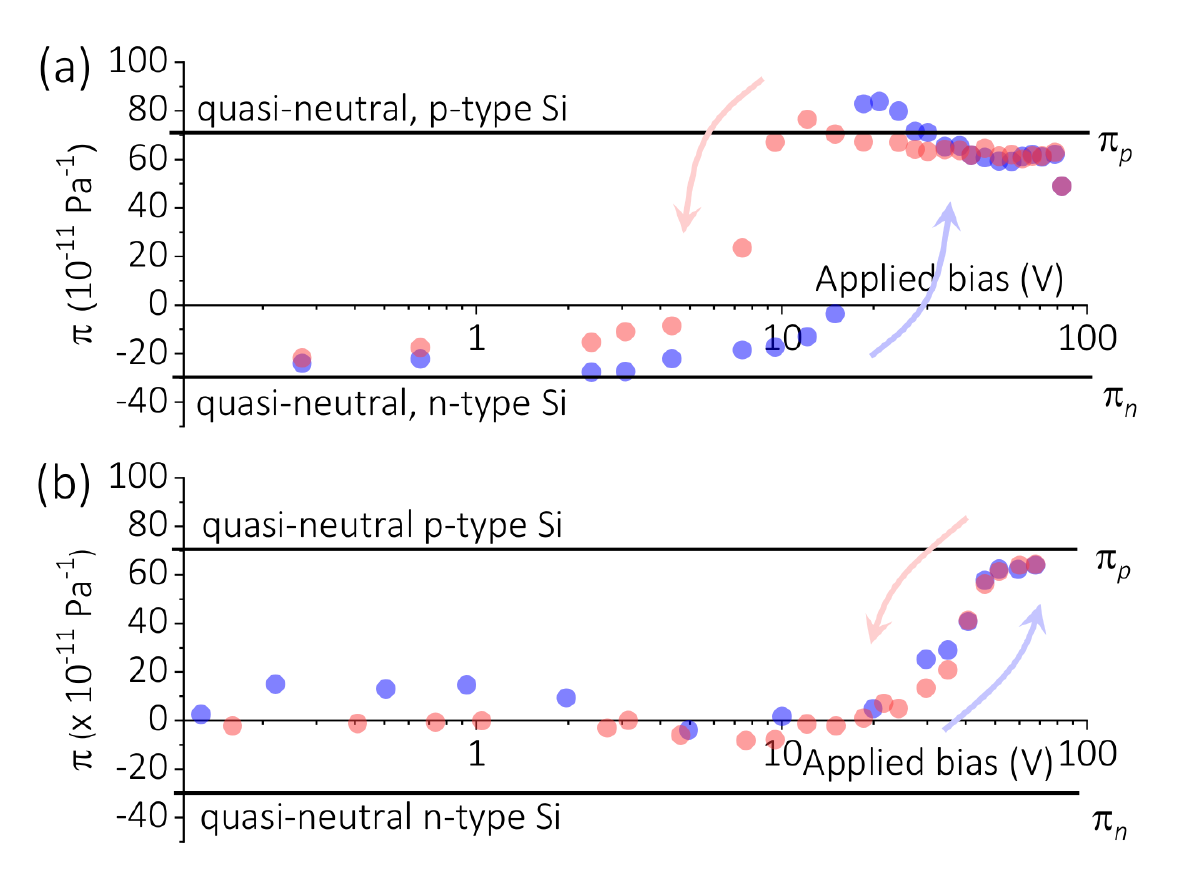}
\caption{PZR $\pi$-coefficient measured under 20 MPa of uni-axial tensile stress parallel to the applied current along the $\langle 110 \rangle$ crystal direction. Up- (blue dots and arrow) and down- (red dots and arrow) sweeps are shown. (a) Results from the defect-engineered sample whose characteristics are shown in Fig.~\ref{data}. Near $V_t$, $\pi$ changes sign and is approximately bounded by the known bulk silicon values \cite{smith1954} given in Eq. (\ref{pip}) and Eq. (\ref{pin}). (b) Results for a device prior to defect engineering. No anomalous (negative) PZR is observed. Note also that the hysteresis only present in the defect-engineered devices.}
\label{PZRdata}
\end{figure}

Figure \ref{PZRdata}(a) shows the PZR $\pi$-coefficient measured simultaneously with the current-voltage characteristic by applying a uni-axial tensile stress of $\approx$ 20~MPa parallel to the current flow along the $\langle110\rangle$ crystal direction. The color code corresponds to the up- and down- sweeps as indicated in Fig.~\ref{data}(a). The $\pi$-coefficient changes sign around the previously defined threshold voltage, $V_t$, varying from approximately $-24 \times 10^{-11}$~Pa$^{-1}$ at low biases to approximately $+65 \times 10^{-11}$~Pa$^{-1}$ at high biases. As Fig.~\ref{PZRdata}(b) indicates, this sign change is \textit{not} observed in the as-prepared devices prior to defect engineering. The threshold voltage at which the switch in sign of the PZR in the defect-engineered devices is observed exhibits the same hysteresis as the current-voltage characteristic in Fig.~\ref{data}(a), but this hysteresis is absent prior to defect engineering (see Fig.~\ref{PZRdata}(b)). The hysteresis is therefore correlated with the presence of defects induced by the Si$^{5+}$ ion implant, as is the anomalous PZR at low bias.

Since the measured PZR switches from approximately that of charge-neutral, n-type silicon given in Eq. (\ref{pin}) at low biases to approximately that of charge-neutral p-type silicon given in Eq. (\ref{pip}) at high biases, and since this switch occurs where the majority carrier type changes from electrons to holes, it is tempting to ascribe the anomalous sign change of the PZR to a simple switch from n-type to p-type PZR. Further analysis however shows that this is incorrect.

\begin{figure}[t]
\includegraphics[clip,width=8 cm] {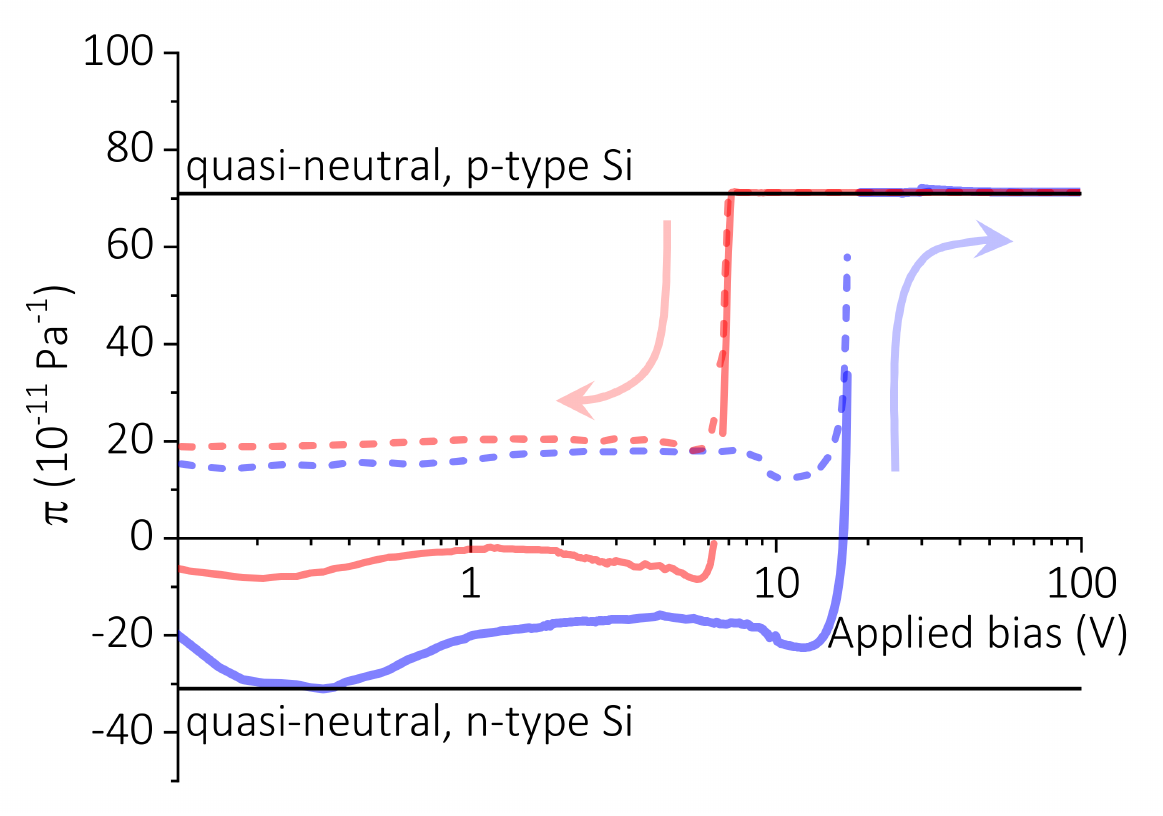}
\caption{PZR calculated using the self-consistent Poisson/drift-diffusion equation solver for the up- and down- voltage sweeps as indicated by the colors and arrows. A stress-induced change in the effective masses alone \cite{smith1954} does not result in anomalous PZR at low bias (dashed lines). Stress-induced shifts in the trap energies consistent with known band deformation potentials does so however (solid lines). The curves show many qualitative similarities with the measured data in Fig.~\ref{PZRdata}(a).}
\label{Calc_PZR}
\end{figure}

Figure \ref{Calc_PZR} shows the calculated PZR with the same color codes for the up- and down- voltage sweeps as used previously. In terms of the origin of the anomalous sign change of the PZR, consideration of the up-sweep curves is instructive. The dashed, blue curve shows the response obtained when only the usual electron and hole mobility changes \cite{smith1954} are accounted for. In this case no anomalous PZR is \textcolor{red}{expected}. The $\pi$-coefficient remains positive, passing from a small value below $V_t$ to the charge-neutral p-type value at high biases where hole injection occurs. This resembles more closely the PZR response obtained prior to defect engineering as shown in Fig.~\ref{PZRdata}(b), suggesting that in the as-processed devices either the density \textcolor{red}{or the stress-dependence of pre-existing trap activation energies} is negligible.

After defect engineering, the electron current below $V_t$ is recombination limited and a stress-dependence of the \textcolor{red}{dominant} SRH \textcolor{red}{process} may then be important. \textcolor{red}{Aside from the effect of stress on the effective mass \cite{smith1954} and hence on the effective densities of states $N_C$ and $N_V$ of the conduction and valence bands,} deformation potential theory \textcolor{red}{suggests that} the most obvious candidate for a stress dependence amongst the \textcolor{red}{SRH} parameters is the trap activation energy, $E_C - E_T$. Trapping cross sections \textcolor{red}{which influence the recombination times $\tau_n$ and $\tau_p$}, are physically related to the shapes of the eigenfunctions, \textcolor{red}{and} are \textcolor{red}{therefore} not expected to significantly change with small applied stresses. \textcolor{red}{Thus in the usual SRH rate expression, \begin{equation} \label{SRH} R_{\textrm{SRH}}=\frac{np - n_i^2}{\tau_n(p+p_1)+\tau_n(n+n_1)}, \end{equation} it is the characteristic electron and hole concentrations, \begin{equation} \label{elecconc} n_1 = N_C \exp\left[-\frac{E_C-E_T}{\textrm{k}_\textrm{B}T}\right] \end{equation} and \begin{equation} \label{holeconc} p_1 = N_V \exp\left[\frac{E_V-E_T}{\textrm{k}_\textrm{B}T}\right], \end{equation} that are the likely origin of the stress dependence of the SRH recombination.}

The solid, blue curve in Fig.~\ref{Calc_PZR} is obtained when $E_C - E_T$ increases by 30~$\mu$V/MPa of tensile stress. The stress-induced increase in $E_C - E_T$ slightly increases the SRH recombination rate resulting in higher currents below $V_t$ and therefore a negative PZR. Above $V_t$ a SCLC hole current proportional to $\mu_p$ which no longer depends on recombination is established, and the PZR naturally tends towards the usual value \cite{smith1954}. This observation also reinforces the conclusion that inter-valley transfer causing velocity saturation is negligible, and that the linear dependence of the characteristic in Fig.~\ref{data}(a) at high voltages is due to an injection barrier \cite{rohr2018}. Similar behavior is observed in the calculated, down-sweep PZR (red curves in Fig.~\ref{Calc_PZR}), where the general form of the curve matches well the measured data in Fig.~\ref{PZRdata}(a), including the hysteresis in $V_t$.

\textcolor{red}{Both the sign and magnitude of the stress-induced change in $E_C - E_T$ estimated here by comparing the numerical model to the transport measurements, are in excellent agreement with independent experimental estimates obtained using a variety of spectroscopic techniques \cite{watkins1965, samara1989, dobaczewski2002}. Using electron spin resonance under uniaxial stress, the singly ionized acceptor form of the divacancy like that which is tentatively identified here shifts by 60~$\mu$V/MPa \cite{watkins1965}, whereas the doubly ionized acceptor form is found to shift under hydrostatic pressure by 12~$\mu$V/MPa using capacitive measurements \cite{samara1989}, and by 51 $\mu$V/MPa using a Laplace DLTS method \cite{dobaczewski2002}. These values are also close to those estimated using first principles calculations \cite{iwata2008}. This agreement gives added weight to the interpretation of the origin of the anomalous PZR that is proposed here.}  

\section{Discussion and Conclusions}

\textcolor{red}{The silicon-on-insulator devices studied here are designed to reproduce the conditions under which giant or anomalous PZR is observed at strains of $\approx$ 0.01 \% in carrier-depleted, nano-silicon objects whose characteristic size is greater than 10 nm \cite{he2006,reck2008,neuzil2010,yang2010,yang2011,kang2012,jang2014,winkler2015}.  As the comparison of the data with the device modeling here shows, background doping levels in the device layer are low enough to ensure that lateral transport is space-charge limited, and the introduction of a nominal $\approx 10^{16}$~cm$^{-3}$ divacancy defect density shortens the SRH lifetimes by several orders of magnitude compared to lightly-doped, defect-free silicon. In comparison, transport through nano-silicon objects of the type exhibiting unusual PZR is likely to be space-charge-limited since in nanowire and nanomembrane geometries surrounded by a low permittivity environment, Gauss' Law dictates that SCLCs are encountered at relatively low voltage thresholds even at relatively high doping densities \cite{grinberg1989,simpkins2008,alagha2017}. Moreover in such objects, high surface-to-volume ratios increase the influence of surface-related SRH recombination resulting in ultra-short recombination times that are comparable to those found here \cite{grumstrup2014}. As such, the results obtained here can be used to partially respond to the questions posed above i.e. why is anomalous PZR only sometimes reported, why is giant PZR so elusive, why is carrier depletion important, and what is the role of electrically active defects?}  

\textcolor{red}{The physical description given here goes beyond the usual descriptive observation that carrier depletion is correlated with unusual PZR \cite{rowe2008}. It explicitly shows that unusual PZR, in this case of anomalous sign, is only obtained for non-equilibrium transport. In the near-equilibrium limit, transport is dominated by equilbrium charge carriers as described by the Drude conductivity and the usual n-type or p-type PZR related to effective mass changes arising from stress-induced valley splittings is obtained \cite{smith1954}. In large nano-silicon objects whose electronic structure is just that of the bulk material, deviation from this PZR is only possible (but not guaranteed) if the non-equilbrium carrier density injected from the contacts is comparable to (or larger than) the equilbrium carrier density. Thus carrier depletion is important for unusual PZR since it ensures that the transport is dominated by non-equilibrium carriers that may be subject to stress-dependent processes other than the usual effective mass changes.}

\textcolor{red}{For non-equilbrium transport this work shows that the PZR depends firstly on whether the carriers injected from the contacts are of the same, or opposite, type to those present at equilibrium. If they are of the same type, a SCLC will be established at some threshold voltage where the characteristic switches from a linear to $V^2$ bias dependence. Since the SCLC is proportional to the carrier mobility \cite{mott1940,grinberg1989} the PZR does not change sign. This is the case for the devices studied here prior to defect engineering (see Fig. \ref{PZRdata}(b)), and is probably also the case in samples where just the usual, bulk PZR in space-charge-limited nano-silicon is found \cite{milne2010}. If the injected carriers are of the opposite type to those present in equilibrium then the characteristic exhibits an initial recombination limited current proportional to $\sqrt{V}$ if the lifetimes are sufficiently short before evolving into a unipolar SCLC at high bias. This is the case for the defect-engineered devices studied here. The PZR then also depends on the stress-dependence of the recombination parameters. If SRH processes are dominant as is usual in silicon, then one immediately understands the influence of the electronically active defects acting as SRH recombination centers in determining the PZR. This is explicitly shown here but it is possible to identify other cases in the literature where anomalous PZR is likely the result of the exact same process \cite{reck2008,jang2014}.}

\textcolor{red}{Finally we come to the question of giant PZR. Can the stress-dependent SRH mechanism result in giant PZR? The answer is probably not, given the magnitude of the stress-induced changes in $E_C - E_T$ estimated here and elsewhere \cite{watkins1965,samara1989,dobaczewski2002}. The order of magnitude of this change($\approx$ tens of $\mu$V/MPa) has also been directly observed for the intrinsic silicon surface defects \cite{li2019b} that limit lifetimes in high surface-to-volume ratio nano-objects \cite{grumstrup2014}. More generally, deformation potentials of this order-of-magnitude are typically observed for any electronic state in an inorganic semiconductor \cite{fischetti1996}. To account for the giant PZR the SRH mechanism described here would therefore require a very peculiar defect with an exceptionally large deformation potential which at this stage seems unlikely. Since current drift is often an issue in transport experiments made in the space-charge-limit, especially in the presence of traps, an alternative explanation for observations of giant PZR is that measurements were unintentionally performed under non-steady-state conditions where charge-trapping-related giant PZR can be observed \cite{li2019}.}

\acknowledgements{The authors acknowledge financial support from the French Agence Nationale de la Recherche (ANR-17-CE24-0005). MF and AT are funded by a grant from the Simons Foundation (601944, MF).}

\bibliography{F:/Publications/References}
\end{document}